\documentclass[numberedappendix]{emulateapj}

\slugcomment{Accepted for publication in AJ}

\shorttitle{The Age of the SMC Cluster NGC\,121}
\shortauthors{Glatt et al.}

\begin{document}

\title{An Accurate Age Determination for the SMC Star Cluster NGC\,121 with HST/ACS
\altaffilmark{*}}

\author{Katharina Glatt\altaffilmark{1,2,3}, 
John S. Gallagher III.\altaffilmark{2}, Eva K. Grebel\altaffilmark{1,3}, 
Antonella Nota\altaffilmark{4}, Elena Sabbi\altaffilmark{4}, 
Marco Sirianni\altaffilmark{4}, Gisella Clementini\altaffilmark{5}, 
Monica Tosi\altaffilmark{5}, Daniel Harbeck\altaffilmark{2},
Andreas Koch\altaffilmark{1,6}, and Misty Cracraft\altaffilmark{4}}
\altaffiltext{*}{Based on observations made with the NASA/ESA Hubble Space Telescope, obtained 
at the Space Telescope Science Institute, which is operated by the Association of Universities 
for Research in Astronomy, Inc., under NASA contract NAS 5-26555. These observations are associated 
with program GO-10396.}
\altaffiltext{1}{Astronomical Institute, Department of Physics and Astronomy, 
University of Basel, Venusstrasse 7, CH-4102 Binningen, Switzerland}
\altaffiltext{2}{Department of Astronomy, University of Wisconsin, 475 North 
Charter Street, Madison, WI 53706-1582}
\altaffiltext{3}{Astronomisches Rechen-Institut, Zentrum f\"ur Astronomie der
Universit\"at Heidelberg, M\"onchhofstr.\ 12--14, D-69120 Heidelberg, Germany}
\altaffiltext{4}{Space Telescope Science Institute, 3700 San Martin Drive, 
Baltimore, MD 21218}
\altaffiltext{5}{INAF - Osservatorio Astronomico di Bologna, Via Ranzani 1, 
40127 Bologna, Italy}
\altaffiltext{6}{Department of Physics and Astronomy, University of California
at Los Angeles, 430 Portola Plaza, Los Angeles, CA 90095-1547}

\begin{abstract}
As first Paper of a series devoted to study the old stellar population
in clusters and fields in the Small Magellanic Cloud, we present deep observations 
of NGC\,121 in the F555W and F814W filters, obtained with the Advanced Camera for 
Surveys on the {\it Hubble Space Telescope}. The resulting color-magnitude diagram 
reaches $\sim 3.5$ mag below the main-sequence
turn-off; deeper than any previous data.  We derive the age
of NGC\,121 using both absolute and relative age-dating methods.  
Fitting isochrones in the ACS photometric system to the observed ridge line of
NGC\,121, gives ages of $11.8 \pm 0.5$~Gyr (Teramo), $11.2 \pm  
0.5$~Gyr (Padova) and $10.5 \pm 0.5$~Gyr (Dartmouth). The cluster ridge line is best
approximated by the $\alpha$-enhanced Dartmouth isochrones.  Placing
our relative ages on an absolute age scale, we find ages of $10.9 \pm 0.5$~Gyr 
(from the magnitude difference between the main-sequence turn-off and the horizontal
branch) and $11.5 \pm 0.5$~Gyr (from the absolute magnitude of the horizontal branch), 
respectively. These five different
age determinations are all lower by 2 -- 3 Gyr than the
ages of the oldest Galactic globular clusters of comparable metallicity.
Therefore we confirm the earlier finding that the oldest globular
cluster in the Small Magellanic Cloud, NGC\,121, is a few Gyr younger than its 
oldest counterparts in
the Milky Way and in other nearby dwarf galaxies such as the Large
Magellanic Cloud, Fornax, and Sagittarius.  If it were accreted into the 
Galactic halo, NGC\,121 would resemble the ``young halo globulars'', although it 
is not as young as 
the youngest globular clusters associated with the Sagittarius dwarf.  The young 
age of NGC\,121 could result from delayed cluster formation in the Small Magellanic Cloud or 
result from the random survival of only one example of an initially small number 
star clusters. 
\end{abstract}


\section{Introduction}

Characterizing old stellar populations provide important constraints on the early
star formation histories of galaxies.
Only the satellite galaxies of the Milky Way (MW) are sufficiently close to
resolve individual stars well below the oldest main-sequence turn-offs, which is a
pre-condition for accurate photometric age dating of old stellar populations.  
All Local Group galaxies, for which adequate data exist, appear to contain stars 
older than 10~Gyr \citep{grebel04}. This result is based on main-sequence
turn-off photometry of globular clusters and field populations in Galactic satellites 
and a few more distant Local Group galaxies \citep[e.g., ][]{brown07, cole07}, as 
well as the detection of horizontal branch stars (including RR Lyrae
variables) in the Local Group and beyond \citep[e.g., ][]
{held00,harb01, saraj02, clementini03, pritzl04}.  

Globular clusters are preferred as the basis for old stellar population 
age tracers since they are usually
single-age, single-metallicity objects facilitating comparative studies.
Moreover, while globular cluster systems exhibit a range of ages 
\citep[e.g., ][]{deang05}, the oldest ones may belong to the 
most ancient surviving stellar systems to  
have completed their formation in the youthful 
Universe \citep[e.g., ][]{moore06}.  In those
nearby galaxies where relative age dating based on main-sequence photometry
was carried out in comparison to the oldest globular clusters in the Milky
Way, no age difference within the measurement accuracy was found
\citep[e.g., ][ and references therein]{grebel04, brown07, cole07}. 
The relative age dating of the oldest identifiable Population~II objects
thus indicates a common epoch of substantial early star formation in the Milky
Way and its companions, although information about a putative, even older
Population III remains to be uncovered in these objects. 

A galaxy that may {\em not}\ share this common epoch of early star
formation -- at least not with respect to its globular clusters
\citep[e.g., ][]{saraj98} -- is the Small Magellanic Cloud
(SMC)\footnote{There may be additional exceptions in more distant dwarf
irregular galaxies regarding the common epoch of earliest Population~II star
formation, although also these galaxies evidently contain old populations
\citep[e.g., ][]{grebel01, maka02}}. The SMC is one of the closest 
and therefore best studied dwarf galaxies orbiting our Galaxy.    

While the SMC hosts a large number of intermediate-age and young star 
clusters, it only contains one 
''old'' globular cluster, NGC\,121, which is also the most massive star
cluster. NGC\,121 is located $\sim$ 2.4\degr 
($\sim$3~kpc) west of the SMC bar at ($\alpha_{J2000.0}$, $\delta_{J2000.0}$) 
= ($0^h26^m47.0^s$, $-71\arcdeg32'12.0''$). 

NGC\,121 is the only cluster in the SMC that is
sufficiently old to have developed an extended red horizontal branch
\citep{stry85} and to contain RR Lyrae stars. Indeed, whether or not to call
a star cluster a globular cluster is a matter of definition.
In this case we refer to \citet{salgir02} who consider Lindsay\,1  
as having a stumpy red clump and not a red horizontal branch. 
Three RR Lyrae stars 
were discovered in NGC\,121 by \citet{thack58}.
\citet{grah75} found a fourth RR Lyrae variable in the cluster and an
additional 75 in a 1 $\times$ 1.3 square degree field centered on NGC\,121.
Studies of various clusters in the LMC and in the MW showed that the
presence of RR Lyrae variables indicates that the parent population is as
old as or older than $\sim$ 10~Gyr. 

An important question is whether NGC\,121 is as old as the typical old
globular clusters in the Large Magellanic Cloud (LMC) and in the MW.
Previous studies found ages ranging from 8 to 14~Gyr for NGC\,121
\citep{stry85, Walker91, migh98, udal98, shara98, dol01} using a variety of
different techniques.  Studies based on the deepest available
color-magnitude diagrams from Hubble Space Telescope (HST) observations
with the Wide Field and Planetary Camera 2 (WFPC2) indicate an age 
of 10 to 10.6~Gyr for NGC\,121, suggesting that this globular cluster 
is several Gyr younger than
the oldest globulars in other nearby galaxies and in the MW 
\citep{shara98, dol01}.

The capabilities of the Advanced Camera for Surveys (ACS) provide an 
improvement in both sensitivity (depth) as well as angular resolution, which is 
essential for a reliable photometric age determination in this dense star cluster.
Here we present deep photometry of NGC\,121 obtained with ACS aboard the 
HST.  We determine the age of NGC\,121
utilizing both absolute and relative methods \citep[e.g., ][]{chab96Sci}.
The current study is the first in a series of papers based on HST studies
of rich intermediate-age and old star clusters in the SMC. 
 
In addition to NGC\,121, six intermediate-age SMC star clusters have been observed 
as part of our program: Lindsay\,1, Kron\,3, NGC\,339, NGC\,416, Lindsay\,38 and NGC\,419. 
We will derive fiducial ridgelines and fit isochrones to obtain accurate ages for each 
cluster using the same reduction techniques and isochrone models as
described here (see $\S$~\ref{sec:obs}-\ref{sec:age}), and will present our results 
in future papers. In Table~\ref{tab:journalobs} 
we list the cluster identification, date of observation, passband, exposure
times and location of all clusters in our HST program (GO-10396; principal investigator: 
J.~S.~Gallagher). 

In the next
Section we describe the data reduction procedure. In $\S$~\ref{sec:CMD} we
present the color-magnitude diagram (CMD) of NGC\,121 and discuss its main
features.  In $\S$~\ref{sec:age} we describe our age derivation methods and
present our results.

\section{Observations and Reductions}
\label{sec:obs}

The SMC cluster NGC\,121 was observed with HST's ACS on 2006 March 21 as
part of our program focused on star clusters and field stellar populations in the SMC. 
The program aims at exploring 
the star formation history and properties of the SMC
using both a number of carefully selected clusters and field regions. 
For NGC\,121 we obtained imaging in the F555W and F814W
filters, which resemble the Johnson V and I filters in their photometric
properties \citep{siri05}.  The images were obtained using the Wide Field
Channel (WFC) of ACS
and cover an area of $200'' \times 200''$  with a pixel scale of
$\sim$ 0.05~arcsec. One set of exposures was taken at the nominal position
of the cluster center. Four long exposures were obtained in each filter for 
hot pixel removal and 
to fill the gap between the two halves of the 4096 $\times$ 4096 pixel detector. 
Each pointing has an exposure time of 496~s in the F555W, and 474~s in the 
F814W filter. Moreover, two short exposures were taken in each filter 
with an exposure time of 10~s in F555W and 20~s in F814W.

\begin{deluxetable*}{ccccccc}
\tabletypesize{\scriptsize}
\tablecolumns{7}
\tablewidth{0pc}
\tablecaption{Journal of Observation}
\tablehead{
\colhead{} & \colhead{Image Name} & \colhead{Date} & \colhead{} & \colhead{Total Exposure Time} & \colhead{R.A.} & \colhead{Dec.} \\
\colhead{Cluster} & \colhead{} & \colhead{(yy$/$mm$/$dd)} & \colhead{Filter} & \colhead{(s)} & \colhead{} & \colhead{} } 
\startdata
NGC\,121&  J96106030 & 2006$/$03$/$21 & F555W & 40.0 & $0^h26^m42.98^s$&$-71\degr32'16.54''$ \\
&J96106040&  &  & 1984.0 & $0^h26^m43.26^s$ & $-71\degr32'14.61''$ \\
&J96106010&  & F814W & 20.0 &  $0^h26^m42.98^s$&$-71\degr32'16.54''$\\
&J96106020&  &  & 1896.0 &  $0^h26^m43.26^s$ & $-71\degr32'14.61''$ \\
Lindsay\,1 &J96105030& 2005$/$08$/$21 & F555W & 40.0 & $0^h03^m53.19^s$ & $-73\arcdeg28'15.74''$ \\
&J96105040&  &  & 1984.0 & $0^h03^m52.66^s$ & $-73\arcdeg28'16.47''$\\
&J96105010&  & F814W & 20.0 & $0^h03^m53.19^s$ & $-73\arcdeg28'15.74''$\\
&J96105020&  &  & 1896.0 & $0^h03^m52.66^s$ & $-73\arcdeg28'16.47''$\\
Kron\,3 &J96107030& 2006$/$01$/$17 & F555W & 40.0 & $0^h24^m41.64^s$ & $-72\arcdeg47'47.49''$\\
&J96107040&  &  & 1984.0 & $0^h24^m41.92^s$ & $-72\arcdeg47'45.49''$\\
&J96107010&  & F814W & 20.0 & $0^h24^m41.64^s$ & $-72\arcdeg47'47.49''$\\
&J96107020&  &  & 1896.0 & $0^h24^m41.92^s$ & $-72\arcdeg47'45.49''$\\
NGC\,339 &J96104030& 2005$/$11$/$28 & F555W & 40.0 & $0^h57^m47.40^s$ & $-74\arcdeg28'26.25''$\\
&J96104040&  &  & 1984.0 & $0^h57^m47.13^s$ & $-74\arcdeg28'24.16''$\\
&J96104010&  & F814W & 20.0 & $0^h57^m47.40^s$ & $-74\arcdeg28'26.25''$\\
&J96104020&  &  & 1896.0 & $0^h57^m47.13^s$ & $-74\arcdeg28'24.16''$\\
NGC\,416 &J96121030& 2006$/$03$/$08 & F555W & 40.0 & $1^h07^m53.59^s$ & $-72\arcdeg21'02.47''$\\
&J96121040&  &  & 1984.0 & $1^h07^m54.09^s$ & $-72\arcdeg21'01.79''$\\
&J96121010&  & F814W & 20.0 & $1^h07^m53.59^s$ & $-72\arcdeg21'02.47''$ \\
&J96121020&  &  & 1896.0 & $1^h07^m54.09^s$ & $-72\arcdeg21'01.79''$\\
Lindsay\,38 &J96102030& 2005$/$08$/$18 & F555W & 40.0 & $0^h48^m57.14^s$ & $-69\arcdeg52'01.766''$\\
&J96102040&  & & 1940.0 & $0^h48^m56.76^s$ &  $-69\arcdeg52'03.07''$\\
&J96102010&  & F814W & 20.0 & $0^h48^m57.14^s$ & $-69\arcdeg52'01.76''$\\
&J96102020&  &  & 1852.0 & $0^h48^m56.76^s$ & $-69\arcdeg52'03.07''$\\
NGC\,419 &J96103030& 2006$/$01$/$05 & F555W & 40.0 & $1^h08^m12.53^s$ & $-72\arcdeg53'17.72''$\\
&J96103040&  &  & 1984.0 & $1^h08^m12.71^s$ & $-72\arcdeg53'15.49''$\\
&J96103010&  & F814W & 20.0 & $1^h08^m12.53^s$ & $-72\arcdeg53'17.72''$\\
&J96103020&  &  & 1896.0 & $1^h08^m12.71^s$ & $-72\arcdeg53'15.49''$\\
\enddata
\label{tab:journalobs}
\end{deluxetable*}

The data set was processed adopting the standard Space Telescope Science
Insitute ACS calibration pipeline (CALACS) to subtract the bias level and
to apply the flat field correction. For each filter, the short and long 
exposures were co-added independently using the MULTIDRIZZLE package
\citep{koek02}.  With this package the cosmic rays and hot pixels were
removed and a correction for geometrical distortion was applied.  The
resulting NGC\,121 data consist of one 40~s and one 1940~s exposure in F555W
and one 20~s as well as one 1896~s exposure in F814W. The two short
exposures allowed us to measure brighter stars that are saturated in the long
exposures. 

The photometric reductions were carried out using the DAOPHOT package in
the IRAF environment \footnote{IRAF is written and supported by the
IRAF programming group at the National Optical Astronomy Observatories
(NOAO) in Tuscon, Arizona. NOAO is operated by the Association of
Universities for Research in Astronomy, Inc. under cooperative agreement
with the National Science Foundation.}. We discarded saturated foreground
stars and background galaxies using the \textit{Source Extractor} package
\citep{bert96}.

Due to the different crowding and signal-to-noise ratio properties of the
long and the short exposure images, photometry involving point spread
function (PSF) fitting was only performed on the long exposures.  For the
short exposures we used aperture photometry, which turned out to yield
smaller formal errors than PSF photometry.  We ran DAOPHOT on our data and
set the detection threshold at 1~$\sigma$ above the local background level
in order to detect even the faintest sources.  The list of stars detected
in the F814W image was then used as coordinate input list to identify the
stars in the F555W image and serve as our coordinate reference. 
49493 sources were found to be
common to both long exposure frames.  For these sources, we performed
aperture photometry using an aperture radius of 3~pixels. We then
constructed a PSF by combining 150 bright and isolated stars that were
distributed fairly uniformlly across the image. Finally, PSF photometry was
carried out.  

The photometric calibration was accomplished by converting the magnitudes
of the individual stars to the standard ACS magnitude system by using an
aperture with a radius of $0.5''$  (or 10~pixels on the image), in combination
with the
aperture correction from the $0.5''$ aperture radius to infinity and the
synthetic zero points for the ACS/WFC \citep{siri05}. The aperture
correction was derived for each frame independently.  The objects found in
both images were cross-identified and merged with a software package
written at the Bologna Observatory by P.\ Montegriffo (private
communication).  Altogether we were able to cover a luminosity range of
$\sim$ 10 magnitudes after combining the resultant photometry of the short
and long exposures.

\begin{figure}
 \epsscale{1}
  \plotone{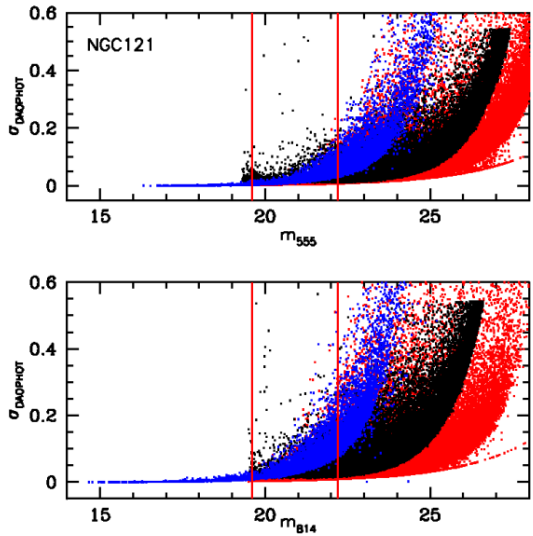}
  \caption{Photometric errors assigned by DAOPHOT to stars in the short exposures
(blue dots), in the aperture photometry from the long (red dots),
and on the PSF photometry from the long (black dots) exposures.  Note the very small
formal errors in the aperture photometry of the short exposures.  Stars
brighter than $\sim 19.3$~mag in the long F555W exposure and brighter than
$\sim 19.5$~mag in the long F814W exposure are saturated and are therefore
not shown.  The lower envelope of the error distribution of the
stars in the short and long F555W exposure (aperture photometry) cross over 
at $m_{555,814}$ = 19.6~mag, and in the aperture and PSF photometry at 
$m_{555,814}$ = 22.2~mag (also indicated by two thin vertical lines).  
Here the photometry
of the short and long exposures was combined.  For the F814W exposures we
chose the same magnitude value in order to avoid introducing a color slope
in the color-magnitude diagram of the resultant data set.}
  \label{fig:NGC121_sigma}
\end{figure}

In Figure~\ref{fig:NGC121_sigma} we show the photometric errors assigned by
DAOPHOT. For stars measured on the short exposures, the formal photometric
errors remain negligible over a wide range of magnitudes.  In the long
exposures, all the brighter stars with $m_{555} < 19.3$~mag and $m_{814} <
19.5$~mag are saturated.  At $m_{555,814} \sim$ 19.6~mag, the short exposure 
(\textit{blue dots}) and aperture photometry from the
long exposure (\textit{black dots}) samples were combined, 
and for stars fainter than $m_{555,814} \sim$ 22.2~mag, long exposure PSF 
photometry (\textit{red dots}) was used. We chose where to 
cut between the aperture and PSF photometry catalogues based on the $m_{555}$ data 
and adopted the same value for $m_{814}$ so as to avoid a color 
slope associated with this division. For our study,
we rejected all stars with a $\sigma$ error larger than 0.2~mag and a
DAOPHOT sharpness parameter $-0.2 \le s \le 0.2 $ in both filters. 
To obtain a superior CMD, we discarded all stars within a radius of $35''$
from the cluster center, which excludes the 
very dense core of the cluster.
With this selection, our final sample contains 17464 stars common in both
filters. 

\section{The Color-Magnitude Diagram}
\label{sec:CMD}

 \begin{figure}
 \epsscale{1}
  \plotone{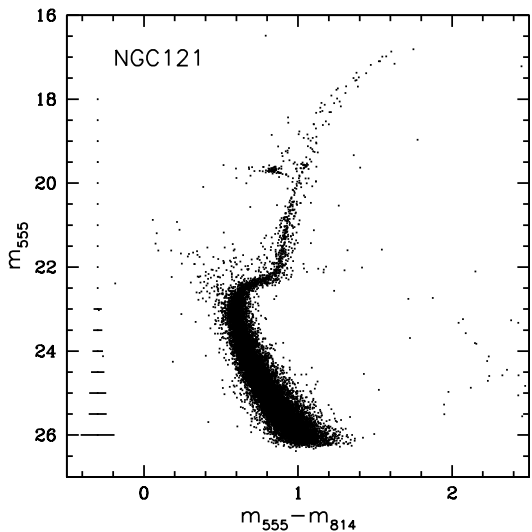}
  \caption{Color-magnitude diagram of NGC\,121 and its surroundings. 
Stars within a radius of 35'' from the cluster center have been discarded.   
All stars with ''good'' photometry ($\sigma \leq 0.2$~mag and $0.2 \geq$ 
sharpness $\geq -0.2$) are shown; 17464 stars in total. Representative 
errorbars (based on the errors assigned by DAOPHOT) are shown on the 
left for the $m_{555}-m_{814}$ color.}
  \label{fig:NGC121_letterfinal}
\end{figure}

The resulting color-magnitude diagram (CMD) of NGC 121 and its
surroundings is shown in Figure~\ref{fig:NGC121_letterfinal}. Our CMD for
NGC\,121 reaches $\sim$3.5~mag below the MSTO ($\sim$~0.5 magnitudes
deeper than the previous deepest available photometry), which allows us to
carry out the most accurate age measurements obtained so far.  The CMD
shows a well-populated main sequence (MS), subgiant branch (SGB), red giant
branch (RGB), horizontal branch (HB), and asymptotic giant branch (AGB).
The gap on the RGB at $m_{555} \sim$ 20~mag is an artificial feature 
due to small number statistics resulting from our exclusion of crowded stars
in the cluster center. 
NGC\,121 appears to be a single-age population object just as one would
expect for a canonical star cluster.  
As expected, there is no obvious 
evidence for field star contamination by younger populations due to the location
of NGC\,121 in a low-density area in the outer parts of the SMC. 
Within the field of view of the ACS and at the high Galactic latitude
of the SMC, Galactic foreground contamination is very low \citep[e.g., ][]{ratna85}.

Another possible contamination source is the massive and extended Galactic 
globular cluster 47\,Tuc, which has a tidal radius of 42.86 arcmin \citep{harris96}
and an angular distance from NGC\,121 of $\sim$ 32~arcmin. 

\begin{figure}
 \epsscale{1}
  \plotone{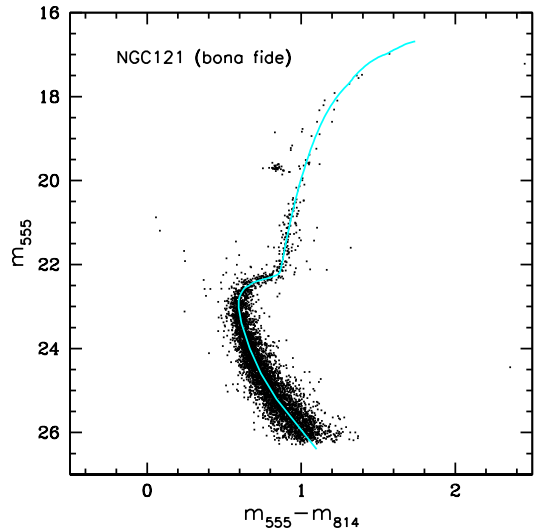}
  \caption{Color-magnitude diagram of all stars within an annulus between $35''$ and $45''$
  of NGC\,121. We used this CMD for 
the determination of a representative color-magnitude ridgeline of NGC\,121
(cyan line). This CMD contains 5112 stars. Only stars with good photometry 
($\sigma \leq 0.2$~mag and $0.2 \geq$ sharpness $\geq -0.2$) are shown.
  }
  \label{fig:NGC121_lettercenter}
\end{figure}

We visually estimated the location of 
the center of NGC\,121 on the image and selected all stars within an annulus of 
$35''$ and $45''$ to create a bona fide sample. This CMD is displayed in 
Figure~\ref{fig:NGC121_lettercenter}.  There is no evidence for a
binary sequence in NGC 121, but we cannot exclude their presence, due to the photometric
error. The aforementioned traces of minor field
contamination have mainly vanished. Due to crowding, incompleteness becomes
significant at the faint end of the MS: This affects particularly faint
stars in the cluster center. 
Hence in Figure~\ref{fig:NGC121_lettercenter} the MS becomes less densely
populated at fainter magnitudes.

The red HB is well populated and extends into the RR Lyrae instability
strip (Clementini et al.\ 2007, in prep).  The presence of a red HB
provides a circumstantial suggestion that NGC\,121 may be younger than old
Galactic and LMC globular clusters, since the HBs of the oldest globular
clusters tend to extend farther into the blue \citep[e.g., ][]{olsz96,
olsen98, mack04}.  Red HBs, however, can also be due to a ''second
parameter'' other than age affecting the HB morphology \citep[e.g.,
][]{lee94, buon97, harb01, catel01}.  Since a true HB is present, an age 
measurement for NGC\,121 can be made using the $\Delta V^{HB}_{TO}$ 
age measurement, which we will do in
$\S$~\ref{sec:relage}.  This method requires the determination of the
apparent mean magnitude of the HB.  Our data yield $m_{555,HB}$ = $19.71
\pm 0.03$~mag for this observable, which is in agreement with the
magnitudes determined by Shara et al.~(1998), Alves \& Sarajedini~(1999)
and Dolphin et al.~(2001).

At $m_{555}$ = $19.58 \pm 0.03$~mag we find the NGC\,121 RGB bump ($m_{555,Bump}$)
which is 0.06~mag brighter than the magnitude found by \citet{alsa99}. The difference
in luminosity is due to the exclusion of the center stars. If we determine the 
$m_{555,Bump}$ on the entire sample, we obtain $m_{555}$ = $19.52 \pm 0.04$~mag,
which is in excellent agreement with the magnitude found by \citet{alsa99}.
This feature is predicted by stellar evolution models, which also show 
that the luminosity of the RGB bump is dependent on the metallicity and age
of the cluster.  When the metallicity is known 
the difference between $V_{HB}$ and $V_{Bump}$ can be used as an age
indicator. 

Above the MS turn-off, \citet{shara98} found 42 candidate blue straggler
stars (BSS). Evolved descendants of the BSSs are important as possible
sources of stars lying above the traditional HB \citep[e.g.,][]{Catelan05}.  
In our ACS study we recovered the \citet{shara98} BSS sample
and also found more stars in the BSS region of which some (about 20) turned out
to be pulsating variables (dwarf Cepheids). These stars will be discussed in 
Clementini et al.\ (2007, in preparation) where we will present the 
results of an HST study of variable stars in NGC\,121.

\section{Age of NGC\,121}
\label{sec:age}
\subsection{Age Based on Isochrone Fits}
\label{sec:isochrone}

\begin{figure}
 \epsscale{1}
  \plotone{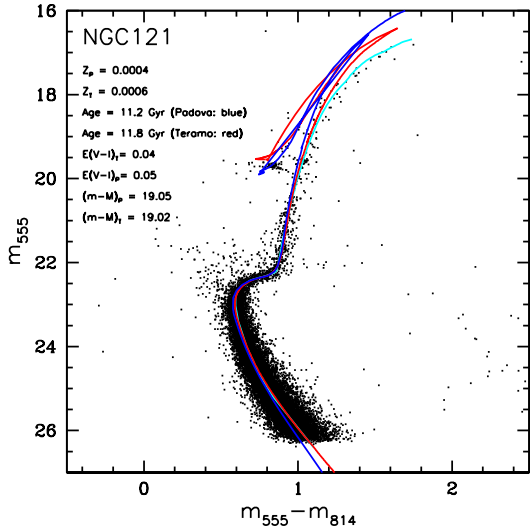}
  \caption{The CMD of NGC\,121 with the best-fitting isochrones of two
different models:  The blue solid line shows the best-fitting Padova
(Girardi, "private communication", Girardi et al. 2000) isochrone that is closest 
to the spectroscopically measured
metallicity of the cluster.  The red solid line is the best-fitting Teramo
\citep{piet04} isochrone approximating the known metallicity.  Neither
model is $\alpha$-enhanced.  The cyan solid line is our fiducial
ridgeline.  The fitting parameters are listed in the plot legend.}
  \label{fig:NGC121_letteriso}
\end{figure}

\begin{figure}
 \epsscale{1}
  \plotone{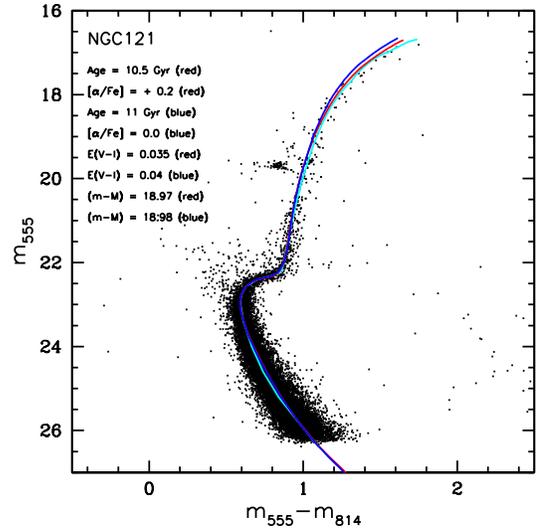}
  \caption{The NGC\,121 CMD with the best-fitting Dartmouth \citep{dotter07} 
isochrones overplotted in red.  As before, the cyan line represents our 
fiducial for NGC\,121. The fit parameters are listed in the plot. Note the
excellent agreement of this $\alpha$-enhanced isochrone with the observed
CMD.}
  \label{fig:NGC121_dart}
\end{figure}

Age determinations of star clusters using isochrones depend crucially on
the interstellar extinction, distance, and metallicity of the cluster, as
well as on the chosen isochrone models. In fitting isochrones to the CMD of
NGC\,121 we adopted the spectroscopic metallicity measurement of [Fe/H] =
$-1.46 \pm 0.10$ from Da Costa $\&$ Hatzidimitriou (1998, see also Johnson
et al.\ 2004) on the metallicity scale introduced by \citet{zinn84} (ZW84).  
The distance and the extinction were treated as free
parameters. The SMC distance modulus is $(m-M)_0$ = $18.88 \pm 0.1$~mag
(60~kpc) \citep[e.g., ][]{Storm04}, but due to the large depth extension of the SMC
along the line of sight we adjusted the distance modulus $(m-M)_0$ to
produce the best isochrone fits to our CMD data. 

For easier comparison to the isochrones, we first derived a fiducial
ridgeline (Table~\ref{tab:ridgeline}), which reproduces the mean location 
of the stellar distribution
in the CMD (exempting the HB).  In order to determine the ridgeline, we
separated the cluster center CMD into three sections: the MS, the SGB and
the RGB. On the MS we determined the mode of the color distribution in
magnitude bins of 0.3 mag width. For the SGB, we performed a linear least
squares fit of a polynomial of 5th order to a Hess diagram of this region
in the CMD.  Finally, the RGB was fit by a third-order polynomial of the
mean color, again in magnitude bins with a size of 0.3 mag each. The
resulting ridgeline is shown in Fig.~\ref{fig:NGC121_lettercenter} as a
cyan line.

We fitted our $m_{555}$ vs.\ $m_{555}$-$m_{814}$ CMD with three different
isochrone models: Padova isochrones (Girardi, "private communication", Girardi et al. 2000)
\footnote{$http://pleiadi.pd.astro.it/isoc\_photsys.02/isoc\_acs\_wfc/index.html$}, 
Teramo isochrones \citep{piet04}, both with scaled solar isochrones
([$\alpha$/Fe] = 0.0), and Dartmouth 
isochrones \citep{dotter07} with both [$\alpha$/Fe] = 0.0 and +0.2. 
The Padova isochrone grid has an age resolution of log(t)=0.05, the Teramo 
isochrone grid of 0.1 Myr and the Dartmouth isochrone grid
of 0.5 Gyr. 
Our adopted spectroscopic metallicity of [Fe/H] = $-1.46$
corresponds most closely to Z = 0.0004 in the Padova models, to Z = 0.0006
in the Teramo models, and to [Fe/H] = $-1.49$ in the Dartmouth models.  All
three sets of isochrone models are available in the standard ACS color
system. 

We fitted a large number of isochrones using different combinations of
reddening, age, and distance.  For each set of models, we selected  the
isochrone that best matched the observed data
(Fig.~\ref{fig:NGC121_letteriso}, Fig.~\ref{fig:NGC121_dart}). 

First we discuss Figure~\ref{fig:NGC121_letteriso}. Our best-fit age using Padova
isochrones is 11.2 Gyr with $(m-M)_0$ = 19.05~mag and $E_{V-I}$ = 0.05. 
The best fitting Teramo isochrone yields an age of t = 11.8~Gyr, $(m-M)_0$ = 
19.02~mag, and $E_{V-I}$ = 0.04. On the MS, both the
Teramo isochrone and the Padova isochrones trace the ridge line almost
perfectly.  At the faint end of the MS, the Padova isochrone continues
further to the blue than the Teramo isochrone and our derived ridge line; however,
this only becomes more apparent at magnitudes of $m_{555}$ $\sim$ 25.5~mag and 
below. Both isochones also provide an excellent approximation to the SGB and to the
lower RGB up to about half a magnitude below the HB.  

At brighter
magnitudes, the two isochrones deviate increasingly to the blue of the
observed upper RGB.  Here the Padova isochrone shows the strongest
difference, deviating by approximately 0.38 mag in color from the observed
tip of the RGB.  The isochrone shows a magnitude for the base of the red HB that is
about 0.5~mag fainter than the observed one. Unlike Teramo and Dartmouth, the Padova
isochrone also models the AGB and its tip, which is $\sim$ 1~mag brighter than the tip
of the RGB.
The Teramo isochrone is too
blue by about 0.23~mag at the magnitude of the tip of the RGB and
indicates a magnitude for the base of the red HB that is 0.2~mag too
bright. 

If we had no prior knowledge of the metallicity of NGC\,121 and were to use
the upper RGB as a metallicity indicator, a better fit would be obtained by
choosing isochrones of a different metallicity or $\alpha$ abundance.  The
problems of various isochrone models of given metallicities in reproducing
the upper red giant branches of globular clusters with the same
metallicities are a well-known problem \citep[e.g., ][]{greb97, greb99}.
Our Figure~\ref{fig:NGC121_letteriso} reflect the general failure of the
chosen stellar evolutionary models to simultaneously reproduce the major
features of CMDs \citep{gall05} in spite of the excellent fit to the lower
RGB, SGB, and MS.  Fortunately the latter 
are the most age-sensitive features of the CMD. 

The isochrone model provided by \citet{dotter07} with [$\alpha$/Fe] = +0.2
yield the best fit to the CMD (Fig.~\ref{fig:NGC121_dart}). The best-fit 
isochrone has the parameters \textit{t} = 10.5~Gyr, $(m-M)_0$ 
= 18.96~mag, and $E_{V-I}$ = 0.035, using the $\alpha$-enhanced isochrones of 
[$\alpha$/Fe] = +0.2. \textit {All} the major features of the
CMD are very well reproduced, including the upper RGB where the isochrone
is offset slightly to the blue relative to the fiducial ridgeline.  This
offset is no more than 0.01 to 0.02 on average along the entire upper RGB;
i.e., even the \textit {slope} of the RGB is very well reproduced along its
entire extent. Unfortunately the stellar evolution models used here terminate at the He
flash, and therefore do not fit the HB or the AGB.

Is our use of $\alpha$-enhanced models justified?
For NGC\,121 a value of ${\rm [Ca/Fe]} = +0.24$ has been measured, which is similar to
the outer LMC cluster Hodge 11 and to the old Galactic outer halo clusters
with ${\rm [Ca/Fe]} = +0.3$ \citep{john04}. Consequently, we
assume that NGC\,121 also is enhanced in $\alpha$-elements. 
We note that in this respect NGC\,121 differs from the general trend
observed in red giant stars in the LMC and in dwarf spheroidal galaxies,
where the [$\alpha$/Fe] ratios at a given [Fe/H] tend to be lower by up to
a few tenths of a dex than in the Galactic halo \citep[e.g., ][]{hill00, 
shet01, fulb02, pritzl05, john06, koch07}.

When we adopt the values for distance and reddening, but fit the cluster with an 
isochrone scaled solar, NGC\,121 gets a slightly older age of 11~Gyr.
The isochrone model with [$\alpha$/Fe] = 0.0 still provides a better fit
than the Teramo or Padova models, but has an offset of $\sim$ 0.05 on average 
along the upper RGB. Past studies found that $\alpha$-enhanced models imply
a higher luminosity and temperature for the same mass than the solar scaled
models and therefore an older age for the same magnitude \citep[e.g., ][]{VandenB00b}. 
The Dartmouth models show exactly the opposite behavior. This is because in these models
an increase in [$\alpha$/Fe] is accompanied by a corresponding increase of the total 
metallicity Z, which makes the isochrones cooler at constant age and [Fe/H]. 

Finally, we note that the derived reddenings agree with the extinction $A_V$ = $0.1 
\pm 0.03$ from the \citet{Schlegel98} maps. The reddening law of \citet{odon94} is assumed. 

\begin{figure}
  \epsscale{1}
  \plotone{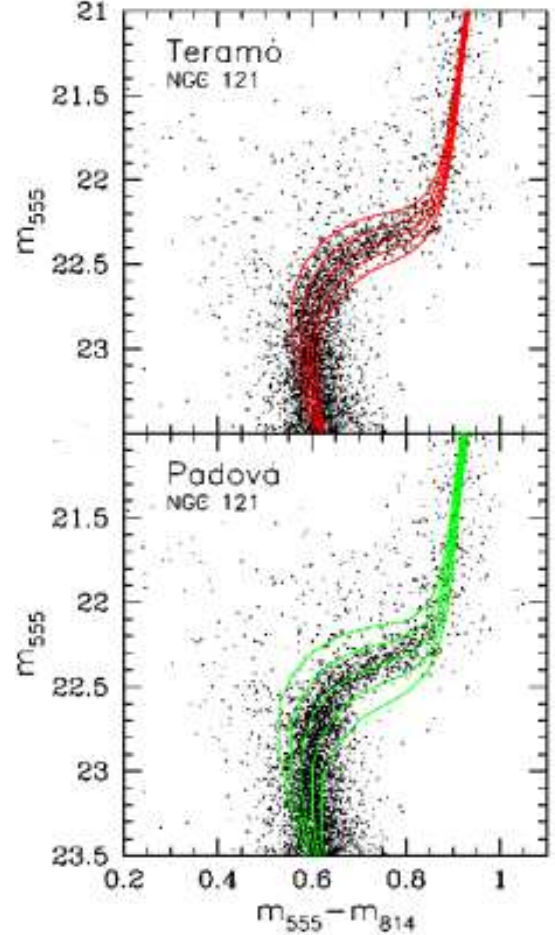}
 \caption{The color-magnitude diagram of NGC\,121 after zooming in on the
region of the main-sequence turn-off, subgiant branch, and lower red giant
branch.  In the upper panel, we show Teramo isochrones as solid lines,
covering an age range of 10, 10.9, 11.8, 12.6, and 13.5~Gyr.  These are 
the age steps in which these isochrones are provided \citep{piet04}.  The 
central isochrone is our chosen best-fitting isochrone.  In the lower panel
we show the same plot for Padova isochrones (solid lines) in the Padova age 
steps of 8.9, 10, 11.2, 12.6, and 14~Gyr (Girardi, "private communication", 
Girardi et al. 2000).  All other
parameters are the same as in Figs.~\ref{fig:NGC121_letteriso} and
\ref{fig:NGC121_dart}.}
 \label{fig:NGC121_isocheck}
\end{figure}

\begin{figure}
  \epsscale{1}
  \plotone{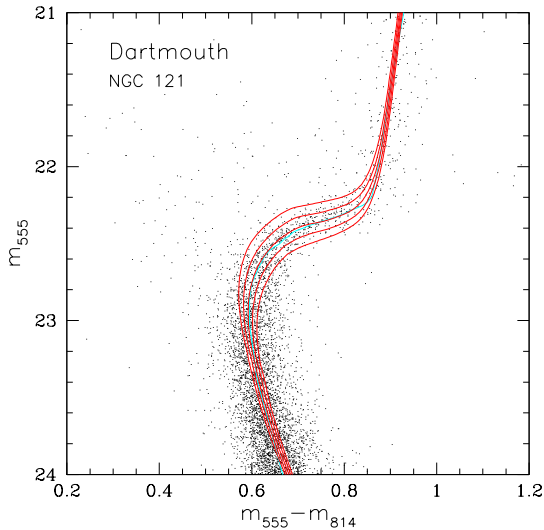}
 \caption{Same as Fig.~\ref{fig:NGC121_isocheck}, but for the
$\alpha$-enhanced Dartmouth isochrones covering an age range of 9.5, 10, 
10.5, 11, and 11.5~Gyr \citep{dotter07}.}
 \label{fig:NGC121_isocheck_dart}
\end{figure}

In the Figures~\ref{fig:NGC121_isocheck} and \ref{fig:NGC121_isocheck_dart}
we show a range of isochrones for the three sets of stellar evolution
models in order to illustrate the age uncertainty in a given model. The
finally chosen, ``best'' isochrone is always displayed along with two
younger and two older isochrones. For the cases of the Teramo and Padova
models, the two isochrones that are one age step younger or older than the
chosen, central isochrone provide an upper or lower envelope for the
location of the high-density part of the SGB, the MS turn-off, and the base
of the RGB.  For the Dartmouth isochrones the
outermost isochrones provide this envelope.  Considering the high quality
of the fit of the central isochrone in this CMD region in all models and
the larger deviations of the adjacent isochrones, we estimate that the
resultant age uncertainty is of the order of approximately $\pm 0.5$ Gyr
for the Teramo and Dartmouth isochrones and $\pm 0.7$ Gyr for the Padova
isochrones.  \\

\begin{deluxetable}{cccc}
\tablecolumns{4}
\tablewidth{0pc}
\tablecaption{Ridgeline of NGC\,121}
\tablehead{
\colhead{$m_{555}-m_{814}$} & \colhead{$m_{555}$} & $m_{555}-m_{814}$ & \colhead{$m_{555}$} \\
\colhead{} & \colhead{} & \colhead{cont.} & \colhead{cont.}}
\startdata
 1.7400&  16.6859& 1.0000&  19.9538 \\
 1.7300&  16.6916& 0.9900&  20.0855 \\
 1.7200&  16.7011& 0.9800&  20.2218 \\
 1.7100&  16.7107& 0.9700&  20.3630 \\
 1.7000&  16.7224& 0.9600&  20.5090 \\
 1.6900&  16.7339& 0.9500&  20.6601 \\
 1.6800&  16.7483& 0.9400&  20.8162 \\
 1.6700&  16.7605& 0.9300&  20.9774 \\
 1.6600&  16.7834& 0.9200&  21.1440 \\
 1.6500&  16.8001& 0.9100&  21.3159 \\
 1.6400&  16.8177& 0.9000&  21.4932 \\
 1.6300&  16.8381& 0.8900&  21.6762 \\
 1.6200&  16.8554& 0.8800&  21.8899 \\
 1.6100&  16.8706& 0.8700&  22.0599 \\
 1.6000&  16.8867& 0.8600&  22.1921 \\
 1.5900&  16.9098& 0.8600&  22.1933 \\
 1.5800&  16.9289& 0.8500&  22.2305 \\
 1.5700&  16.9540& 0.8400&  22.2520 \\
 1.5600&  16.9712& 0.8300&  22.2660 \\
 1.5500&  16.9874& 0.8200&  22.2770 \\
 1.5400&  17.0008& 0.8100&  22.2890 \\
 1.5300&  17.0153& 0.8000&  22.3050 \\
 1.5200&  17.0309& 0.7900&  22.3180 \\
 1.5100&  17.0477& 0.7800&  22.3280 \\
 1.5000&  17.0658& 0.7700&  22.3380 \\
 1.4900&  17.0851& 0.7600&  22.3480 \\
 1.4800&  17.1057& 0.7500&  22.3570 \\
 1.4700&  17.1276& 0.7400&  22.3670 \\
 1.4600&  17.1508& 0.7300&  22.3770 \\
 1.4500&  17.1753& 0.7200&  22.3880 \\
 1.4400&  17.2013& 0.7100&  22.4010 \\
 1.4300&  17.2286& 0.7000&  22.4120 \\
 1.4200&  17.2575& 0.6900&  22.4260 \\
 1.4100&  17.2877& 0.6800&  22.4520 \\
 1.4000&  17.3195& 0.6700&  22.4710 \\
 1.3900&  17.3528& 0.6600&  22.5020 \\
 1.3800&  17.3877& 0.6500&  22.5190 \\
 1.3700&  17.4241& 0.6400&  22.5410 \\
 1.3600&  17.4621& 0.6300&  22.5720 \\
 1.3500&  17.5018& 0.6200&  22.6240 \\
 1.3400&  17.5432& 0.6100&  22.6890 \\
 1.3300&  17.5962& 0.6000&  22.7610 \\
 1.3200&  17.6378& 0.5950&  22.8500 \\
 1.3100&  17.6904& 0.5950&  23.0310 \\
 1.3000&  17.7246& 0.6125&  23.4000 \\
 1.2900&  17.7605& 0.6670&  24.0000 \\
 1.2800&  17.7983& 0.7400&  24.6000 \\
 1.2700&  17.8379& 0.8400&  25.2000 \\
 1.2600&  17.8794& 0.9700&  25.8000 \\
 1.2500&  17.9231& 1.1000&  26.4000 \\
 1.2400&  17.9690&& \\
 1.2300&  18.0171&& \\
 1.2200&  18.0676&& \\
 1.2100&  18.1205&& \\
 1.2000&  18.1761&& \\
 1.1900&  18.2343&& \\
 1.1800&  18.2952&& \\
 1.1700&  18.3590&& \\
 1.1600&  18.4257&& \\
 1.1500&  18.4955&& \\
 1.1400&  18.5685&& \\
 1.1300&  18.6447&& \\
 1.1200&  18.7242&& \\
 1.1100&  18.8071&& \\
 1.1000&  18.8936&& \\
 1.0900&  18.9837&& \\
 1.0800&  19.0775&& \\
 1.0700&  19.1752&& \\
 1.0600&  19.2767&& \\
 1.0500&  19.3823&& \\
 1.0400&  19.4920&& \\
 1.0300&  19.6059&& \\
 1.0200&  19.7241&& \\
 1.0100&  19.8367&& \\
\enddata
\label{tab:ridgeline}
\end{deluxetable}

\subsection{Empirical Age Estimates}
\label{sec:relage}

\subsubsection{Vertical Method}

To check the reliability of the isochrone ages, we use a
reddening-independent method to derive relative ages of NGC\,121. This
method is also independent of the photometric zeropoint of our data.  This
''vertical method'' relies on the fact that the absolute
magnitude of the MSTO depends on the age of the cluster \citep[e.g., ][]{alsa99}, 
while the absolute
magnitude of the HB remains approximately age-independent for clusters
older than $t \gtrsim$ 10~Gyr \citep[e.g., ][]{girar01}. We measure the
apparent magnitudes of the MSTO and HB, i.e., $V_{TO}$ and $V_{HB}$, in
order to obtain the magnitude difference $\Delta V^{HB}_{TO}$. With
increasing age, a cluster has generally larger values of this parameter
since the MS moves to fainter magnitudes.  Unfortunately, the determination
of these two points comes with significant uncertainties.  $\Delta
V^{HB}_{TO}$ is hard to measure accurately both because of the width of the HB in
luminosity and the MSTO's vertical extent in the turn-off region. 


First, we calculate the magnitude difference $\Delta m^{HB}_{TO,555}$
between the HB and the MSTO, as originally described by \citet{Iben68}.
Because $m_{555}$ is proportional to V and we are only interested in the
magnitude difference, which we measure at constant color, 
there is no need to transform the magnitudes from the
ACS system to {\textit V} and {\textit I} magnitudes \citep{siri05}. We
follow the general definition of the MSTO as the bluest point along the MS;
in our case represented by the bluest point on the ridgeline. We find the
MSTO at $m_{TO,555}$ = $22.98 \pm 0.05$~mag, $(m_{555}-m_{814})$ = $0.59
\pm 0.005$~mag.  As described in Section~\ref{sec:CMD}, the mean HB
magnitude is $m_{HB,555}$ = $19.71 \pm 0.03$~mag.  This yields a $\Delta
m_{TO,555}^{HB} = 3.27 \pm 0.06$~mag. Our result is only slightly lower
than former values, e.g., $\Delta V^{HB}_{TO}$ =  3.32 determined by
\citet{stry85}, 3.33 by \citet{shara98} or 3.29 by \citet{dol01}.
\citet{buon89} published a mean value of $\Delta V^{HB}_{TO} = 3.55$~mag
for old Galactic halo globular clusters (GC). The $\Delta m^{HB}_{TO,555}$
of NGC\,121 is 0.28~mag smaller than this value, which indicates that NGC\,121
is younger than most of the older \textit{Galactic} GCs.

\citet{walker92} (see also Buonanno et al. 1989) found a relation between age and $\Delta V^{HB}_{TO}$
based on a study of 41 Galactic globular clusters. Adopting again a
metallicity of [Fe/H] = $-1.46 \pm 0.10$, we find with the formula $\log t
= -0.045{\rm [Fe/H]} + 0.37 \Delta V^{HB}_{TO} - 0.24$ an age of $10.9 \pm
0.5$~Gyr for NGC\,121.  

\subsubsection{Age Estimate using $M_V(HB)$}

In order to compare our data directly with \citet{dol01}, we use the age
calibration provided by \citet{chab96b} as was done by these authors. 
For this purpose we must adopt the metallicity Fe/H] = $-1.19 \pm 0.12$
of \citet{daco98} on the metallicity scale introduced by \citet{cargra97} (CG97) 
to be consistent with \citet{dol01}'s calculation. Note that elsewhere in the paper we are
using metallicities on the ZW84 scale which agrees reasonably well with the
spectroscopically derived chemical abundances \citep{john04}. Using this method
and the higher metallicity we determine an age of $9.7 \pm 1.0$~Gyr for NGC\,121, 
which is similar to the
age that \citet{dol01} found. We have to emphasize that this age is 
younger than the absolute ages determined in Section~\ref{sec:isochrone} due to the
different metallicity scale. If we use our preferred ZW84 metallicity scale with this
method we obtain an older age of $10.8 \pm 1.0$~Gyr.

While the measurement of $V_{TO}$ is affected by significant observational
errors ($\sim$ 0.05~mag), the color of the MSTO is well-defined.  \citet{chab96a}
found that the usage of a point on the SBG brighter than the MSTO and 0.05
mag redder ($V_{BTO}$) provides more precise relative ages than $V_{TO}$.
\citet{chab96a} provide a conversion between $M_{V}(BTO)$ and $V$, $I$ data
for a grid of five metallicities.  We choose the conversion for [Fe/H] $=
-1.5$ because it is closest to the metallicity of NGC\,121.  In our data, we
measured $m_{BTO,555}$ = $22.45 \pm 0.02$ at $(m_{555}-m_{814})_{BTO}$
= $0.64 \pm 0.005$~mag.  To convert $m_{BTO,555}$ to the absolute magnitude
$M_{BTO,555}$ we use the distance modulus derived above. This yields
$M_{BTO,555} = 3.49 \pm 0.1$~mag. With the modified calibration by
\citet{john99} we obtain an age of $11.50 \pm 0.5$~Gyr for NGC\,121. We have
summarized all our age results in Table ~\ref{tab:ages}.  


\subsubsection{Red Bump}

Calculating $\Delta m_{555,Bump}^{HB}$ = $m_{555,Bump}-m_{555,HB}$ we find
$-0.13 \pm 0.05$~mag. There is a general trend of increasing RGB bump
brightness with decreasing age \citep{alsa99}, assuming that a different
''second parameter'' is not affecting the position of the RGB bump.
\citet{alsa99} presented a $V_{Bump}^{HB}$ vs.\ [Fe/H] diagram (their Fig.\
6), where the brightness difference between the RGB bump and the HB is
plotted against the cluster metallicity.  If we assume that the HB
magnitude does not critically depend on age, then NGC\,121 is slightly older
than 10~Gyr based on this relation.

These comparative results are sensitive to the abundance of
$\alpha$-elements in NGC\,121 as compared to Galactic globular clusters.
Many nearby Galactic globular clusters are enhanced in $\alpha$-elements
relative to the solar value \citep{john04}. As mentioned earlier, spectroscopic
results for NGC\,121 indicate that this cluster is similarly enhanced in $\alpha$
elements, which means that the relative ages should not be affected as long as we 
confine ourselves to the comparison of globular clusters with similar $\alpha$-element 
ratios. The fact that NGC\,121 does not follow the trend of reduced [$\alpha$/Fe] ratios
observed in other nearby dwarf galaxies facilitates both our relative age
determinations.   

\begin{deluxetable}{ccc}
\tablecolumns{3}
\tablewidth{0pc}
\tablecaption{Ages for NGC\,121 derived in this paper}
\tablenote{All derived ages are listed along with the method applied. In all cases,
we adopted the ZW84 metallicity scale.}
\tablehead{
\colhead{Age [Gyr]} & \colhead{Method} & \colhead{Reference for method} }
\startdata
$10.9 \pm 0.5$ & $\Delta V^{HB}_{TO}$ &\citet{walker92} (ZW) \\ 
$11.5 \pm 0.5$ & $M_{V}(BTO)$	      &\citet{chab96a} \\
$10.8 \pm 1.0$ & $M_V(HB)$	      &\citet{chab96b}  \\
$11.8 \pm 0.5$ & Isochrones	      &\citet{piet04}	\\ 
$11.2 \pm 0.7$ & Isochrones	      &Girardi, "private communication", \\
&&					\citet{girar00}    \\ 
$10.5 \pm 0.5$ & Isochrones	      &\citet{dotter07} \\ 
\enddata
\label{tab:ages}
\end{deluxetable}

\subsubsection{Relative Age of NGC\,121}

In Table~\ref{tab:comp_ages_ZW} we compare the relative age of NGC\,121 with
those for a sample of Galactic globular clusters. While this comparison sample is 
located in the Galactic halo, some objects may have formed outside the Galaxy and 
might have been subsequently captured or accreted. We list the clusters by their 
identification in column (1). The [Fe/H] values are given in column (2) in the scale
by \citet{zinn84}. Column (3) shows the $\Delta V_{TO}^{HB}$ and column (4) the ages
obtained by using the \citet{walker92} calibration. Finally, column (5) gives the
relative age difference of these clusters to NGC\,121 $\delta (t)_{W}$. For 
$\Delta V_{TO}^{HB}$ we adopted
the values from \citet{deang05}, unless differently stated
(see footnotes of Table~\ref{tab:comp_ages_ZW}). 

The clusters are listed in order
of increasing $\Delta V_{TO}^{HB}$ and are divided into two
groups.  The first group shows nine ''pure'' GCs with similar metallicities
as NGC\,121.  Among these nine clusters is NGC\,2808 for which multiple
MSTOs have been found \citep{piotto07}.  Its $\Delta V_{TO}^{HB}$ value,
derived prior to the study by \citep{piotto07}, is comparatively small.
Even though those nine clusters have all similar metallicities, the spread
in $\Delta V_{TO}^{HB}$ and therefore in age is quite large: NGC\,1262 shows
the lowest $\Delta V_{TO}^{HB}$ = 3.24 and is similar in age to NGC\,121,
while NGC\,6656 has $\Delta V_{TO}^{HB}$ = 3.55, which makes it $\sim$3 Gyr
older than NGC\,121. 

The second group includes a subset of Galactic halo clusters that appear to be 
significantly younger than the average of the Galactic globular cluster population
\citep[e.g., ][]{Rosenberg99, VandenB00a, salar02}. Some members of this
group are listed in the last part of Table~\ref{tab:comp_ages_ZW}.  NGC\,362
and NGC\,288 are known to be a second parameter cluster pair of different
ages, as reflected in their different $\Delta V_{TO}^{HB}$ \citep[e.g.,
][]{fupe96, catelan01, bella01}. As NGC\,362 has a similar $\Delta
V_{TO}^{HB}$ and [Fe/H] as NGC\,121, it should therefore be of a similarly
young age.  
Other members of this group of young halo globulars are IC~4499
\citep{ferraro95}, Ruprecht~106 \citep[assumed to be 3--5 Gyr younger than the
bulk of the Galactic globulars with similar metallicities; ][]{daco92,
buon93}, Arp~2 \citep{buon95a}, Terzan 7 \citep{buon95b}, and Pal~14 \citep{saraj97}.

\begin{deluxetable*}{ccccc}
\tablecolumns{5}
\tablewidth{0pc}
\tablecaption{Comparison of globular clusters ages (vertical method)}
\tablenote{Ages were determined using the \citet{walker92} calibration.
The metallicity of NGC\,121 was adopted from \citet{daco98}. The other results for
NGC\,121 were derived in this Paper.  The data for NGC\,6656 where taken from 
\citet{Rosenberg99}; 
for IGC~4499 from \citet{ferraro95}; for Pal~12 from \citet{rutl97} and for
Pal~14 from \citet{ferraro95} and \citet{ferraro95,saraj97}.  All other values 
were taken from \citet{deang05}.}
\tablehead{
\colhead{Cluster} & \colhead{$[Fe/H]_{ZW84}$} & \colhead{$\Delta V_{TO}^{HB}$} & 
\colhead{Age [Gyr]} &\colhead{$\delta(t)_{W}$} }
\startdata 
NGC\,121  &-1.46 &3.27 & 10.9 &0      \\
NGC\,1261 &-1.32 &3.24 & 10.4 &-0.5  \\
NGC\,5272 &-1.66 &3.24 & 10.8 &0.1   \\
NGC\,2808 &-1.36 &3.25 & 10.6 &-0.3  \\
NGC\,3201 &-1.53 &3.28 & 11.0 &0.1   \\
NGC\,5904 &-1.38 &3.34 & 11.4 &0.5   \\
NGC\,6254 &-1.55 &3.37 & 11.9 &1.0   \\
NGC\,6218 &-1.40 &3.48 & 12.9 &2.0   \\
NGC\,6752 &-1.54 &3.53 & 13.7 &2.8   \\
NGC\,6656 &-1.41 &3.55 & 13.7 &2.8   \\
\hline
Pal~12   &-0.94 &3.17 &  9.45 &-1.3   \\
IG~4499   &-1.75 &3.25 & 11   &0.1    \\
NGC\,362  &-1.33 &3.27 & 10.7 &-0.2  \\
Pal~14    &-1.65 &3.33 & 11.7 &0.8    \\ 
NGC\,288  &-1.40 &3.45 & 12.6 &1.7    \\
\enddata
\label{tab:comp_ages_ZW}
\end{deluxetable*}

Among the theories that try to explain the existence of these young objects
is the model according to which they are intergalactic clusters captured by
the MW \citep{buon95b}, or clusters formed during interactions
between the MW and the Magellanic Clouds assuming that they are on bound
orbits \citep{fupe95}.
\citet{zinn93} argued that the apparent young halo globular clusters formed
in dwarf galaxies that later merged with the MW.  Hence the Galactic
globular clusters are assumed to be a mixture of objects that formed with the
MW itself (old halo group) and others accreted from destroyed dwarf
satellites (young halo clusters) (see also Mackey \& Gilmore 2004).  
At least six globular clusters are believed to be associated with the
Sagittarius dwarf galaxy \citep[e.g., ][ and references therein]{carraro07}, 
providing support for the accretion scenario.  Note,
however, that Sagittarius is contributing both old (M54, Ter~8, Arp~2) and
``young'' (Ter~7, Pal~12, Whiting~1) globular clusters to the MW.
Similarly, the only other Galactic dSph galaxy known to contain globular
clusters, Fornax, would contribute both kinds of globulars
\citep{buon98,buon99} if it were to merge with our Galaxy.  This also holds
for the LMC \citep[][see also discussion in Grebel, Gallagher, \& Harbeck 2003]{olsen98}. 

If we take all the ages determined in this 
paper into account, we find that NGC\,121 is consistently 2--3~Gyr younger than the 
oldest Galactic globular clusters \citep[absolute age $\sim$ 13~Gyr according to ][]
{Kraus03} and LMC globular clusters.
The age offset remains when comparing NGC\,121 to old Galactic
globular clusters in the same metallicity range (see Tab.~\ref{tab:comp_ages_ZW},
upper panel). We also show that NGC\,121 is not as young as the youngest 
Galactic and Sagittarius globular clusters, some of which are $\sim$ 2~Gyr younger than
NGC\,121.

\section{Summary and Discussion}

We derived ages for the old SMC globular cluster NGC\,121 based on our high
dynamic range HST/ACS
photometry that extends at least three magnitudes below its MSTO.  In order
to obtain absolute ages, we applied three different isochrone models.
These isochrone models yielded ages of $11.2 \pm 0.7$~Gyr (Padova), $11.8
\pm 0.5$~Gyr (Teramo), and $10.5 \pm 0.5$~Gyr (Dartmouth).  We find the
$\alpha$-enhanced Dartmouth isochrones provide the closest approximation
to the MS, SGB, and RGB, whereas the other models cannot reproduce the
slope of the upper RGB. High-resolution spectroscopy indicates that NGC\,121
is indeed $\alpha$-enhanced \citep{john04}, a property that it shares with many 
of the old outer Galactic halo globulars. Given the proximity of NGC\,121 to the 
SMC on the sky and its distance, its physical association with the SMC seems 
well-established. 

Our determinations of relative ages for NGC\,121 are consistent with the results of 
our absolute age determination. Relative age estimates, when converted to an 
absolute age scale, are $10.9 \pm 0.5$~Gyr ($\Delta V_{TO}^{HB}$), $10.8 \pm 1.0$ 
($M_V(HB)$) and $11.5 \pm 0.5$~Gyr ($M_{V(BTO)}$). These numbers agree well 
with the absolute age derivations. Our results confirm that NGC\,121 is 2--3~Gyr 
younger than the oldest MW and LMC clusters (as also found in earlier WFPC2 
studies).

NGC\,121 is similar in age to the youngest globular cluster
in the Fornax dSph \citep{buon99}, and to several of the young Galactic halo clusters. 
On the other hand, NGC\,121 is not as young as some of the Sgr dwarf galaxy's globular
clusters or the youngest Galactic globular clusters.

It is intriguing that the SMC -- in
contrast to other Galactic companion dwarf galaxies with globulars -- does not 
contain any old classical globular clusters. But given the existence of only one cluster and the 
question of star cluster survival, this could be a result of the one survivor 
from the SMC's epoch of globular cluster formation randomly 
sampling an initial distribution of star cluster ages.  On the other hand, in low-mass galaxies
without  bulges, spiral density waves, and shear it is much more difficult to destroy globular
clusters through external effects. That this cluster is both 
younger than the Galactic mean and enhanced in $\alpha$-elements may 
have interesting implications for the early development of the SMC.

It also is intriguing that the only globular cluster in the
SMC is not very metal-poor.  The SMC must have experienced
substantial enrichment prior to the formation of NGC\,121.  In the LMC, where
two main epochs of the formation of populous compact star clusters
have been found \citep[e.g., ][]{Bertelli92}, a few globular
clusters are found that are old enough to exhibit blue HBs.  
Interestingly, these globular clusters, which are similarly old
as the oldest Galactic globulars \citep{olsen98}, have 
a similar metallicity to NGC 121 \citep{john04} (e.g., NGC\,1898,
NGC\,2019), indicating very early chemical
enrichment.  The MW also contains old classical globular
clusters (with blue HBs) that have similarly high metallicities
as the somewhat younger NGC 121. Evidently, the conditions for
and the efficiency of star formation varied in
these three galaxies at early epochs.

After NGC 121 formed there was a hiatus in surviving stars clusters 
and thus possibly in cluster formation
activity in the SMC: The second oldest SMC cluster is Lindsay 1 with 
an age of $\sim$ 8~Gyr (Glatt et al. 2007, in preparation).
Since then compact populous star clusters formed fairly  
continuously until the present day in the SMC (e.g., Da Costa
2002) -- in contrast to both the LMC and the MW.  
In forthcoming papers on our ACS photometry of SMC clusters
and field populations we will explore the evolutionary history
of the SMC in more detail.  Clearly, clues about the early
star formation history of the SMC will have to come from
its old field populations.

\acknowledgments 
We thank the anonymous referee for extremely useful suggestions to improve our paper.
We gratefully acknowledge support by the Swiss National Science Foundation through grant 
number 200020-105260 and 200020-113697. Support for program GO-10396 was provided 
by NASA through a grant from the Space Telescope Science Institute, which is operated 
by the Association of Universities for Research in Astronomy, Inc., under NASA contract 
NAS 5-26555. We warmly thank Paolo Montegriffo to
provide his software and Leo Girardi for the Padova isochrones in the standard ACS color 
system. Gisella Clementini and Monica Tosi have been partially supported by PRIN-MIUR-2004 
and PRIN-INAF-2005, and Jay Gallagher also obtained helpful additional support from the 
University of Wisconsin Graduate School.

\clearpage
\end{document}